\documentclass[11pt]{article}
\usepackage{epsfig}
\newcommand{\bara}{\begin{array}{c}}
\newcommand{\eara}{\end{array}}
\newcommand{\as}{\alpha_s}
\newcommand{\asb}{\bar{\alpha_s}}
\newcommand{\asm}{\alpha_s(\mu^2)}

\newcommand{\Sud}{{\cal S}}
\newcommand{\sg}{\sigma}
\newcommand{\sgo}{\sigma_0}
\newcommand{\bmu}{\bar{\mu}}

\def\pt{p_{\mbox{\tiny T}}} 
\def\pts{p_{\mbox{\tiny T}*}}

\def\tfnp{\tilde{F}^{NP}(\pt)}
\def\ta{\tilde{a}}
\def\ptlim{p_{\mbox{\tiny T\,lim}}}
\def\MSbar{\overline{\rm MS}}
\def\aone{A^{(1)}}
\def\atwo{A^{(2)}}
\def\bone{B^{(1)}}
\def\btwo{B^{(2)}}

\def\prt{perturbative }
\def\nprt{non-perturbative }
\def\nn{\nonumber}
\def\rar{\rightarrow}

\def\lapproxeq{{\ \lower 0.6ex \hbox{$\buildrel<\over\sim$}\ }}
\def\gapproxeq{{\ \lower 0.6ex \hbox{$\buildrel>\over\sim$}\ }}

\def\pl#1#2#3{
        {\it Phys.\ Lett.\ }{\bf #1} (#2) #3}

\def\prl#1#2#3{
        {\it Phys.\ Rev.\ Lett.\ }{\bf #1} (#2) #3}

\def\prep#1#2#3{
        {\it Phys.\ Rep.\ }{\bf #1} (#2) #3}
\def\pr#1#2#3{
        {\it Phys.\ Rev.\ }{\bf #1} (#2) #3}
\def\np#1#2#3{
        {\it Nucl.\ Phys.\ }{\bf #1} (#2) #3}
\def\epj#1#2#3{
        {\it Eur.\ Phys.\ J.\ }{\bf #1} (#2) #3}

\def\jhep#1#2#3{
        {\it JHEP }{\bf #1} (#2) #3}

\begin{document}  
\vspace*{-2cm}  
\renewcommand{\thefootnote}{\fnsymbol{footnote}}  
\begin{flushright}  
hep-ph/0103089\\
BNL-HET-01/8\\
IPPP/01/10\\
DCPT/01/20\\
March 2001\\  
\end{flushright}  
\vskip 65pt  
\begin{center}  
{\Large \bf Soft Gluon Resummation in Transverse
Momentum Space for Electroweak Boson Production at Hadron Colliders}\\
\vspace{1.2cm} 
{\bf  
Anna~Kulesza${}^1$\footnote{Anna.Kulesza@bnl.gov} and  
W.~James~Stirling${}^{2}$\footnote{W.J.Stirling@durham.ac.uk}  
}\\  
\vspace{10pt}  
{\sf 1) Department of Physics, Brookhaven National Laboratory, Upton, NY
11973, U.S.A. \\  
  
2) Institute for Particle Physics Phenomenology, University of Durham,  
Durham DH1 3LE, U.K.}
  
\vspace{70pt}  
\begin{abstract}
The distribution of $W$ and $Z$ bosons produced with small transverse momentum ($\pt$) at hadron 
colliders receives important contributions from large logarithms arising from soft gluon emission. Although 
conventionally the all-orders resummation of  these  `Sudakov' logarithms is performed in impact parameter (Fourier transform) space,   $\pt$-space resummation is also possible, and offers certain advantages.
We present a detailed phenomenological analysis of  $W$ and $Z$ production at small $\pt$
 at the Tevatron $p \bar p$ collider, using $\pt$-space resummation. A good description of the CDF and D0 data
can be obtained provided a significant non-perturbative contribution is included. We also present 
predictions for the LHC.
\end{abstract}
\end{center}  
\vskip12pt

\setcounter{footnote}{0}  
\renewcommand{\thefootnote}{\arabic{footnote}}  
\vfill  
\clearpage  
\setcounter{page}{1}  
\pagestyle{plain} 


\section{Introduction}

The description of gauge boson production at hadron colliders has recently
attracted much theoretical interest, especially in the light of future
high precision experiments at the Tevatron and the LHC~\cite{LHC,LesHouches,Tevatron}.
Reliable predictions can only be obtained if soft gluon radiation effects are 
correctly taken into account.  Theoretically the soft
gluon emission manifests itself in the presence of large logarithmic
corrections (Sudakov logarithms). For the particular case of the 
transverse momentum ($\pt$) distribution of a boson produced with invariant
mass $Q$, the Sudakov logarithms are the logarithms of the ratio $Q^2/\pt^2$.  
In the small $\pt$ limit the logarithms diverge and the standard fixed-order
perturbation theory approach breaks down. However, a finite result can be
recovered if the soft gluon emission is accounted for to all orders in $\as$.
This is achieved by resumming the logarithmic corrections. 

Resummation can be performed either directly in transverse momentum ($\pt$) space
or in  the Fourier conjugate impact parameter ($b$) space. The most leading logarithmic
contributions, of the form $\as^n \ln^{2n-1} (Q^2/\pt^2)$, can be directly resummed
in $\pt$ space (the so-called Double Leading Logarithm
Approximation)~\cite{DDT}. The impact parameter ($b$) space
method~\cite{CSS} allows one to resum subleading logarithms, including those
`kinematic' logarithms arising as a direct result of transverse momentum
 conservation~\cite{PP}.  
Although very successful theoretically, the $b$ space method suffers from
certain deficiencies and drawbacks which need to be `fixed' in order to obtain
a satisfactory agreement between the theoretical predictions and 
experimental data. For example, one experiences difficulties 
when matching the resummed (small $\pt$) and fixed-order
(large $\pt$) predictions. Moreover, it is impossible to make predictions for {\em any}
value of $\pt$ without a prescription dealing with the \nprt
regime of large $b$\footnote{
Recently it has been argued~\cite{QZ} that the $b$ space drawbacks
discussed here are less relevant at Tevatron and LHC energies.}. These
difficulties can be naturally circumvented if the resummation is performed  in $\pt$
space~\cite{EV}. 
Unfortunately the $\pt$ space methods which have been
developed so far for resumming subleading logarithms are derived from the $b$ space
approach~\cite{EV,FNR,KS}, and as such they simply provide an approximation to the $b$ space
result. However, the goal is to 
develop a phenomenologically useful $\pt$ space expression that reproduces {\em all}
the good features of the $b$ space resummation without the drawbacks related
to this method. 

In a previous paper~\cite{KS} we proposed a $\pt$ space resummation
formalism at the parton level which resums the first four `towers' of
logarithms (i.e. terms of the form $\as^n \ln^{2n-m}(Q^2/\pt^2),$ $m=1,.. ,4$),
{\em including} the effects of transverse momentum conservation.
The differences between our formalism (KS) and other $\pt$ space 
approaches (FNR~\cite{FNR} and EV~\cite{EV})  were discussed in~\cite{KS2}. 
Here we want to concentrate on the practical applications of our formalism, 
in particular on the comparison with available Tevatron data.

In our analysis we use the most recent sets of CDF~\cite{CDF:Z} and D0
data~\cite{D0:Z} on $Z$ production and D0 data~\cite{D0:W} on $W$
production. Since we are only interested here in the resummed part of the cross
section and do not perform matching with the fixed-order part, we do not consider data 
above $\pt>25$ GeV. In this $\pt$ range the resummed part accounts for almost the entire
cross section. 

In a manner similar to the $b$ space formalism, the $\pt$ space formalism
is incomplete without a prescription for dealing with the \nprt effects. Indeed previous phenomenological analyses have shown that the very
small $\pt$ region is dominated by  \nprt contributions. Here we use the method of 
introducing \nprt effects in  $\pt$
space first proposed in~\cite{EV}. We investigate the form and size of the \nprt
contributions obtained from fits to the data. 
We finish our investigations by commenting on $W$ and $Z$ boson
production at the LHC.

\section{Theoretical cross section for $p \bar{p} \rar W,Z +X$}
\label{Theo:Xsec}

In this section we summarise the derivation of the main 
theoretical results for $\pt$ space used in fits to the data.
For the discussion of the $b$ space results the reader is referred to~\cite{CSS,ERV}.

The resummed part of the theoretical cross
section in  $\pt$ space for a Drell-Yan-type process follows from the $b$ space formula,
cf.~\cite{CSS}
\begin{eqnarray}
{d \sg \over d \pt^2\, d Q^2} &=& {\sg_0 \over Q^2} \sum_{q} e_q^2 
\int^1_0 \,d x_A \, d x_B \, \delta \left(x_A x_B - {Q^2 \over s}\right)
\times \nn \\
&& {1 \over 2} \int_0^{\infty}  db \, b \,J_0( \pt b) \, \exp[{\cal S}(b,Q)]
\,\tilde{f}^{\prime}_{q/A}\left(x_A, {b_0 \over b}\right) 
\,\tilde{f}^{\prime}_{\bar{q}/B}\left(x_B, {b_0\over b}\right)\,. \nn \\
\end{eqnarray}
with $\sigma_0 = 4 \pi \alpha^2 / (9 s)$, $b_0=2\exp(-\gamma_E)$, and  where
\begin{eqnarray}
S(b,Q^2) = - \int_{b_0^2 \over b^2}^{Q^2} \frac{d\bar\mu^2}{\bar\mu^2} 
\bigg[ \ln \bigg ( {Q^2\over\bar \mu^2} \bigg ) A(\alpha_S(\bar\mu^2)) +
B(\alpha_S(\bar\mu^2)) \bigg ] \,,\label{Sbs} \\
A(\alpha_S) = \sum^\infty_{i=1} \left(\frac{\alpha_S}{2 \pi} \right)^i A^{(i)}\
, \quad
B(\alpha_S) = \sum^\infty_{i=1} \left(\frac{\alpha_S}{2 \pi} \right)^i
B^{(i)}\,.
\label{AB}
\end{eqnarray}

We first consider  the non-singlet (NS) cross-section, i.e. we introduce
\begin{displaymath} 
\tilde{f}^{\prime}_{q/H} = f^{\prime}_{q/H} - f^{\prime}_{\bar{q}/H} \nn \\
\end{displaymath}
as  modified higher-order NS parton distributions. The modified parton distributions are 
related to the $\MSbar$ parton distributions, $f$,
by a convolution~\cite{CSS,DS,ERV}
\begin{equation}
f^\prime_{a/H} (x_A,\mu) = \sum_{c}
\int_{x_A}^{1} {d z \over z} \,
C_{ac}\left( {x_A \over z},\mu \right)
f_{c/H} \left( z, \mu \right)\ ,
\end{equation}
where ($a,b \neq g$)
\begin{eqnarray*}
C_{ab}(z,\mu) &=&
\delta_{ab} \Bigg\{\delta(1-z)
+\asb(\mu) C_F  \left[ 1-z
+\left( \frac{\pi^2}{2}-4 \right) \delta(1-z)\right] \Bigg\} \ , \\
C_{ag}(z,\mu) &=& \asb(\mu) T_R
\Big[ 2 z( 1-z) \Big]\,,
\end{eqnarray*}
and $\asb(\mu) = \frac{\as(\mu)}{2 \pi}$, $C_F=4/3$, $T_R=1/2$.

The $N$-th moment of the cross section with respect to
$\tau = Q^2/s$ has the form
\begin{eqnarray}
{\cal M}(N) &=& \int d\tau\, \tau^N {Q^2 \over \sgo}
{ d \sg \over d \pt^2 \, dQ^2} \nn \\
&=&
\sum_{q} e_q^2 {1\over 2}
\int_{0}^{\infty} db \, b  \, J_{0}(\pt b)\, \exp{[{\Sud(b,Q)}]}
\,\tilde{f}'_{q/A}(N, {b_0 \over b})
\,\tilde{f}'_{\bar{q}/B}(N, {b_0 \over b}) \ .
\end{eqnarray}
Solving the DGLAP equation for the $N$-th moment of the modified 
parton distribution  ${\displaystyle \tilde{f}'_{q/H}(N,Q)= \int_0^1 d x_H x_H^N
  \tilde{f}'_{q/H}(x_H,Q)}$, and integration by parts lead to (cf.~\cite{EV})
\begin{eqnarray*}
{\cal M}(N) &=& 
\frac{d}{d\pt^2} \Bigg\{
\sum_{q} e_q^2 \; \tilde{f}^\prime_{q/A}(N,\pt)
 \tilde{f}^\prime_{\bar{q}/B}(N,\pt) \nn \\
&& \times \int_{0}^{\infty} dx \, J_{1}(x)\,
\exp \left[\Sud(x,Q) - 2 \int_{{b_0^2\pt^2 \over x^2}}^{\pt^2}
{d \bmu^2\over \bmu^2} \gamma^\prime_N(\asb(\bmu))
\right] \Bigg\}\nn \,,
\end{eqnarray*}
where $x=\pt b$.

In order to obtain an expression for the hadron level cross 
section, the following approximation is introduced 
\begin{equation}
\exp \left[\Sud(x,Q) - 2 \int_{{b_0^2\pt^2 \over x^2}}^{\pt^2}
{d \bmu^2\over \bmu^2} \gamma^\prime_N(\asb(\bmu))
\right] \approx \exp \left[\Sud(x,Q)\right] \,.
\label{anomdim:approx}
\end{equation}

The above equality is exact for the first four towers of logarithms; it is
only the fifth tower that contains the first modified anomalous dimension
coefficient $\gamma^{\prime (1)}_N$. This can be easily seen by expanding the
exponential in~(\ref{anomdim:approx}) (assuming here a fixed coupling constant
for simplicity)
\begin{eqnarray*}
\lefteqn{\exp \left[\Sud(x,Q) - 2 \int_{{b_0^2\pt^2 \over x^2}}^{\pt^2}
{d \bmu^2\over \bmu^2} \gamma^\prime_N(\asb(\bmu))
\right]=} \nn \\
&{\displaystyle  \sum_{N=0}^{\infty}{(-1)^N \over N!}}& \left[
{ 1 \over 2 }( \aone \asb  + \atwo\asb^2 +...) (L + L_b)^2 
+ (\bone\asb +\btwo \asb^2+...) (L + L_b) \right. \nn \\
&& \left.+ 2 (\gamma^{\prime (1)}_N\asb +\gamma^{\prime (2)}_N \asb^2+ ...) L_b
\right]^N \,,
\end{eqnarray*}    
where $L=\ln(Q^2/\pt^2)$ and $L_b=\ln(x^2/b_0^2)$. 
The first term containing $\gamma^{\prime (1)}_N$ which does not vanish after
integration over $x$ is of the form 
\mbox{$\asb^N {\aone}^{N-1}
  \gamma^{\prime (1)}_N L^{2(N-2)} L_b^3$}. The same statement holds also for the singlet parton
distribution functions.

The resulting expression 
\begin{eqnarray}
{\cal M}(N) = \frac{d}{d\pt^2} \Bigg\{
\sum_{q} e_q^2 \; \tilde{f}^\prime_{q/A}(N,\pt)
\tilde{f}^\prime_{\bar{q}/B}(N,\pt) 
\int_{0}^{\infty} dx \, J_{1}(x)\,
\exp [\Sud(x,Q)] \Bigg\}
\end{eqnarray}
can now be transformed back to momentum space by the means of the inverse Mellin transform
\begin{eqnarray}
{d \sg \over d \pt^2\, d Q^2} &=& {\sgo \over Q^2} \sum_{q} e_q^2 
\int^1_0 \,d x_A \, d x_B \, \delta \left(x_A x_B - {Q^2 \over s}\right) \times \nn \\
&& \frac{d}{d\pt^2} \Bigg\{\int_{0}^{\infty} dx \, J_{1}(x)\,
\exp [\Sud(x,Q)]
\,\tilde{f}^{\prime}_{q/A} (x_A, \pt ) 
\,\tilde{f}^{\prime}_{\bar{q}/B} (x_B, \pt)\Bigg\} \,.
\end{eqnarray} 

At the parton level we calculated the quantity\footnote{see Eq.~(21) in~\cite{KS}}
\begin{eqnarray}
{1 \over \sgo}{d \sg \over d \pt^2} &=& 
 -\frac{1}{2 \pt^2} \int_{0}^{\infty} dx \,x\, J_1(x)\frac{d}{d x}\exp [\Sud(x,Q)] \nn\\
&=& -\frac{1}{2 \pt^2} \exp( \Sud_\eta (Q)) \int_{0}^{\infty} dx \,x\,
J_1(x)\frac{d}{d x} \exp[\tilde{\Sud}(x,Q)] \nn \\
&=& {\asm \aone \over 2 \pt^2 \pi} 
e^{ \Sud_{\eta}} \sum_{N=1}^{\infty} 
\Bigg({-\asm \aone \over \pi}\Bigg)^{N-1}
\sum_{m=0}^{N-1} {1 \over m!} 
\sum_{k=0}^{N-m-1} {1 \over k!} \nn
\\
&\times&
\sum_{l=0}^{N-m-k-1} {1 \over l!} 
\sum_{j=0}^{N-m-k-l-1} {1 \over j!}
\sum_{i=0}^{N-m-k-l-j-1} {1 \over i! (N-m-k-l-j-i-1)!} \nn \\
&\times&
c_2^m c_3^k  c_4^l c_5^j c_6^i  c_1^{N-m-k-l-j-i-1} 
\sum_{n=1}^{6} n c_n \tau_{N+m+2k+3l+4j+5i+n-2}\, \nn \\
&\equiv& -\frac{1}{2 \pt^2}  \Sigma_1(\pt,Q) \,,
\label{ptspace:parton}
\end{eqnarray}
in terms of resummed towers of logarithms in $\pt$ space. Here
$\tilde{\Sud}(x,Q)=\Sud(x,Q)-\Sud_\eta (Q)$ with $\Sud_\eta$ and $c$
coefficients defined in~\cite{KS}.  
The expression for ${\displaystyle \int_{0}^{\infty} dx J_1(x)\exp[\Sud(x,Q)]}$ can be
derived in a similar manner 
\begin{eqnarray}
&&\hspace{-1cm}\int_{0}^{\infty} dx J_1(x)\exp[\Sud(x,Q)]=\exp(\Sud_{\eta}) \sum_{N=1}^{\infty} \Bigg({-\alpha_S(\mu^2) A^{(1)}\over
  \pi}\Bigg)^{N-1}
\sum_{m=0}^{N-1} {1 \over m!} 
\sum_{k=0}^{N-m-1} {1 \over k!}\nn
\\
&\times&
\sum_{l=0}^{N-m-k-1} {1 \over l!} 
\sum_{j=0}^{N-m-k-l-1} {1 \over j!}
\sum_{i=0}^{N-m-k-l-j-1}{1 \over i! (N-m-k-l-j-i-1)!} \nn \\
&\times&
c_2^m c_3^k  c_4^l c_5^j c_6^i  c_1^{N-m-k-l-j-i-1} 
\tau_{N+m+2k+3l+4j+5i-1} \nn \\
&\equiv&  \Sigma_2(\pt,Q) \,,
\label{int:hadron}
\end{eqnarray}
Finally we arrive at the $\pt$ space formula for the Drell-Yan cross section
at the hadron level
\begin{eqnarray}
{d \sg \over d \pt^2\, d Q^2} &=& {\sgo \over Q^2} \sum_{q} e_q^2 
\int^1_0 \,d x_A \, d x_B \, \delta \left(x_A x_B - {Q^2 \over s}\right) \times \nn \\
&& \frac{d}{d\pt^2} \left\{ \Sigma_2(\pt,Q)
\,f^{\prime}_{q/A} (x_A, \pt ) 
\,f^{\prime}_{\bar{q}/B} (x_B, \pt)  \right\} \,.
\label{ptspace:hadron}
\end{eqnarray}
In what follows we will refer to the result in  Eq.~(\ref{ptspace:hadron}) as the KS
hadron-level formula in $\pt$ space.

In principle, the parton level formula~(\ref{ptspace:parton}) allows us to
 resum {\em any} number of towers of logarithms. In practice, however, the fifth tower of
 logarithms cannot be fully taken into account due to the lack of knowledge
 of the coefficient $A^{(3)}$.
Since our approximation~(\ref{anomdim:approx}) is valid only up to the fifth
 tower too, Eq.~(\ref{ptspace:hadron}) can be used to resum the first four
 towers of logarithms, in other words the summation in~(\ref{int:hadron}) stops
at $N=4$. In  \cite{KS}, the contributions from fifth and higher towers
were estimated to be numerically very small in the region of $\pt$  of interest.

The analogous expression for the transverse momentum distribution of a massive vector boson 
$V$ produced in $p \bar p \rar V+X$ is
\begin{eqnarray}
{d \sg \over d \pt} &=& \sgo \sum_{q q'}  U^V_{q q'} 
\int_0^1 \,d x_A \, d x_B \, \delta \left(x_A x_B - {M_V^2 \over s}\right)
\times \nn \\
&& \frac{d}{d\pt} \left\{ \Sigma_2(\pt,M_V)
\,f^{\prime}_{q/A} (x_A, \pt ) 
\,f^{\prime}_{q'/B} (x_B, \pt)\right\} \,,
\label{KS:hadron}
\end{eqnarray}
where 
\begin{eqnarray} 
\sgo &=& \frac{\pi \sqrt{2} G_F}{N} \nn \\
U^V_{q q'} &=&  \left\{ \bara |V_{q q'}|^2  \qquad \qquad \qquad  \quad V =
W^{\pm} \,,\\
\!\!\!\! (V_q^2 +A_q^2)\delta_{q q'} \qquad \qquad  \!\!\!\! V = Z\,,  \eara \right. 
\end{eqnarray} 
where $V_{q q'}$ denotes the appropriate CKM matrix element, and  $V_q$,
$A_q$ are the vector and axial couplings of the $Z$ boson to quarks.

In practice it is convenient to split the differentiation in~(\ref{KS:hadron})
into two terms 
\begin{eqnarray}
{d \sg \over d \pt} &=& \sgo \sum_{q q'}  U^V_{q q'} 
\int_0^1 \,d x_A \, d x_B \, \delta \left(x_A x_B - {M_V^2 \over s}\right)
\times \nn \\
&& \left\{ 
-{1 \over \pt}  \Sigma_1(\pt,M_V) 
f^{\prime}_{q/A} (x_A, \pt ) \, f^{\prime}_{q'/B} (x_B, \pt) 
\right.\nn \\
&& + \left. \Sigma_2(\pt,M_V)
\frac{d}{d\pt } \left[ f^{\prime}_{q/A} (x_A, \pt ) 
\,f^{\prime}_{q'/B} (x_B, \pt) \right] \right\} \,.
\label{KS:hadron2}
\end{eqnarray}
This trick allows us to apply an inevitable numerical
derivative only to the product of the parton distributions and  not
to the whole expression, leading to a reduction of the numerical error.

\subsection{Inclusion of the non-perturbative effects in $\pt$ \mbox{space}}

The form of the  \nprt ansatz in $\pt$ space is expected to be important
only in regions where perturbation theory fails,
i.e. at the very low values of $\pt \leq 2-3$ GeV. In contrast, 
the higher $\pt$ region can be described purely by
the resummed perturbative QCD expression. 
We choose to incorporate the low energy effects using the form 
of the $\pt$ space \nprt function $\tfnp$ advocated in~\cite{EV}
\begin{equation}
\tfnp = 1-\exp{[-\tilde{a}\, \pt^2]}\,. 
\label{F:NP:pt}
\end{equation}
The role of this function is to account for the distribution in the very low
$\pt$ region, and here we are assuming that the shape is approximately gaussian.
  However in order to combine this with the perturbative result, the latter needs to be `frozen'  or `switched off' at some critical value of $\pt$ where the coupling $\as$ becomes large. 
A similar freezing is required in the $b$ space approach where the coupling is 
effectively $\as(1/b)$. In other words we require not only  (i) a form 
$\tfnp$ for the distribution in the
\nprt region, but also (ii) a prescription for moving smoothly
from the perturbative to the \nprt region.  
One possibility for the latter is the 
 `freezing' prescription of~\cite{EV} 
\begin{equation}
\pts = \sqrt {\pt^2 + \ptlim^2 \exp{\left[-\frac{\pt^2}{\ptlim^2}\right]}}\,
\label{ptstar}
\end{equation}
 which has the property
\begin{equation}
\pts =  \left\{ \bara  \pt \, ,   \qquad \qquad \qquad  \quad \pt \gg \ptlim \, , \\
                        \ptlim  \, ,     \qquad \qquad\quad  \quad    \pt \ll \ptlim   \, .  \eara \right. 
\end{equation}
It is important to note that there are {\em two} pieces of information contained
in this definition: the value of the limiting value $\ptlim$ and the abruptness of the transition
 to this value. The use of a gaussian function in the definition (\ref{ptstar}), compared to say
a power law function, implies a rapid transition that, as we shall see below, is consistent
with the data.

Applying the above prescription to our expression~(\ref{KS:hadron}) leads to
\begin{eqnarray}
{ d \sg \over d \pt} &=& \sgo \sum_{q q'}  U^V_{q q'} 
\int_0^1 \,d x_A \, d x_B \, \delta \left(x_A x_B - {M_V^2 \over s}\right)
\times \nn \\
&& \Bigg\{ 
-{1 \over \pts}  {d \pts \over d \pt} \Sigma_1(\pts,M_V)
f^\prime_{q/A}(x_A,\pts) \,f^\prime_{q'/B}(x_B,\pts) \tfnp
 \nn \\
&& + \Sigma_2(\pts,M_V) {d \pts \over d \pt} {d \over d \pts} \left[ 
f^\prime_{q/A}(x_A,\pts) \,f^\prime_{q'/B}(x_B,\pts) \right] \tfnp
\nn \\
&& + \Sigma_2(\pts,M_V) f^\prime_{q/A}(x_A,\pts) \,f^\prime_{q'/B}(x_B,\pts)
{d \over d \pt} \tfnp \Bigg\} \,.
\label{KS:FNP}
\end{eqnarray} 

Note that  the simple form of $\tfnp$ in the present
framework does not take into account a possible dependence 
on $Q$ and $x$. This is in contrast to the $b$ space treatment of~\cite{LY},
where an $x$--dependent linear term in $b$ was added to the argument of the gaussian
\nprt function. It has also been argued \cite{CSS} that the width of the 
\nprt gaussian distribution, in our case the parameter $1/\tilde{a}$, should increase
linearly with $\log Q$~\footnote{
Indeed, fits of a \nprt gaussian distribution to the low energy
data~\cite{ESWbook} typically give much larger values of $\tilde{a}$ ($\approx 0.55$~GeV$^{-2}$),
which  suggests a strong dependence of $\ta$ on $Q$.}.
Since in the present case we are only interested in 
$W,Z$ production at a single collider energy, the values of $Q$ and $x$
are essentially fixed at $M_V$ and $M_V/\sqrt{s}$ respectively. Therefore
we are not able to say anything about the form of the dependence of the \nprt parameters
on $Q$ and $x$. Nor will we investigate different functional forms for
$\tfnp$ --- the simple gaussian form in (\ref{F:NP:pt}) will allow perfectly good fits
to the Tevatron data.

The lack of information on the $x$ dependence of the \nprt contributions should be borne
in mind when considering the predictions for $W$ and $Z$ production at the LHC.

\section{Results and discussion}
 
For the parton level cross section we have
advocated~\cite{KS} the use of the renormalization scale $\mu_R=\pt^{(2/3)} Q^{(1/3)}$
as a means of eliminating certain logarithmic terms from the Sudakov factor
and thus increasing the reliability of our approach. Since the renormalization scale
determines the strength of the coupling in the theoretical predictions, it
must somehow depend on the size of the transverse momentum and we require the
choice of the scale to reflect this fact. Moreover, for values of
$\pt$ where perturbative QCD can be safely applied and for the values of $Q$
considered here, such  a $\mu_R$ is always bigger then the $b$ quark mass, 
thus lessening the relevance of the correction due to the treatment of
 quark mass thresholds.  Another obvious choice
for the renormalization scale is $\mu_R=\pt$.%
\footnote{Other choices of $\mu_R$ have been considered in the
  literature, for example the authors of~\cite{EMP} proposed to take
  $\mu_R=\sqrt{\pt^2+Q^2}$.}
However, since we find only a very small dependence of the resummed part of
the cross section on the choice of $\mu_R$, from now on we
 use $\mu_R=\pt^{(2/3)} Q^{(1/3)}$ as the default choice for the KS approach.

The Drell-Yan cross section~(\ref{ptspace:hadron}) has been derived in the limit of
a fixed number of quark flavours,
$N_f$, which implies that no quark mass threshold effects are considered. In
the original $b$ space approach, the dependence on $N_f$ enters in the Sudakov 
factor through the $\atwo,\ \btwo$ coefficients and through the $\beta$ function in the
expansion of $\as$. For our $\pt$ space method we propose to change $N_f$ according to
the number of flavours active at the renormalization
scale at which $\as$ is calculated while the remainder of the expression is derived in the
massless quark limit. With the choices of
the scale we use this roughly corresponds to the energy scale of the emitted 
gluons and fits comfortably into the physical picture of the process. 
Changing the number of active quark flavours $N_f$ as $\pt$ varies immediately leads to the
problem of obtaining reliable predictions free of unphysical
discontinuities. To overcome this we use an analytically
extended $\as^{\MSbar}$ scheme which incorporates  finite-mass
quark threshold effects into the running of the coupling, as proposed by
Brodsky {\it et al.}~\cite{BGMR}. By connecting the
coupling directly to the analytic and physically-defined effective charge 
scheme, the authors of~\cite{BGMR} obtain an analytic expression for the effective 
number of flavours which is a continuous function of the renormalization
scale and the quark masses. 

\vspace{1cm} 
The results presented below are for the Tevatron experiments, CDF and D0, at 
$\sqrt{s}=1.8$~TeV. Unless stated otherwise we use the factorization scale $\mu_{f}=\pt$, MRST99 parton distribution functions~\cite{MRST99} (central
gluon), branching ratios $BR(Z \rar e^-e^+)=3.366\%$, $BR(W \rar e \nu)=11.1\%$ 
and the world average value of the strong coupling $\as(M_Z)=0.1175$. To normalize the 
theory predictions to the data we take only those experimental points
with $\pt<15$~GeV.

There is a significant amount of Tevatron data on $W$ and $Z$ production
that should, in principle, allow a precise determination of the  \nprt parameters from 
fits to the data. However since the measurement of the $W$ transverse 
momentum  requires correcting for
detector effects that are much stronger that in the $Z$ measurement
case, for the purpose of this analysis we take only the $Z$ data. 
Again, we consider only those experimental points with $\pt<15$ GeV for the fit range. 
The overall normalization is taken as a free parameter, since we are primarily
concerned with the {\em shape} of the distributions.~\footnote{The 
{\em total} $W$ and $Z$ cross sections
are known to be well described by NNLO perturbation theory, 
see e.g. \cite{MRST99}.}

\begin{figure}[h]
\begin{center}
\vspace{-2.5cm}
\mbox{\epsfig{figure=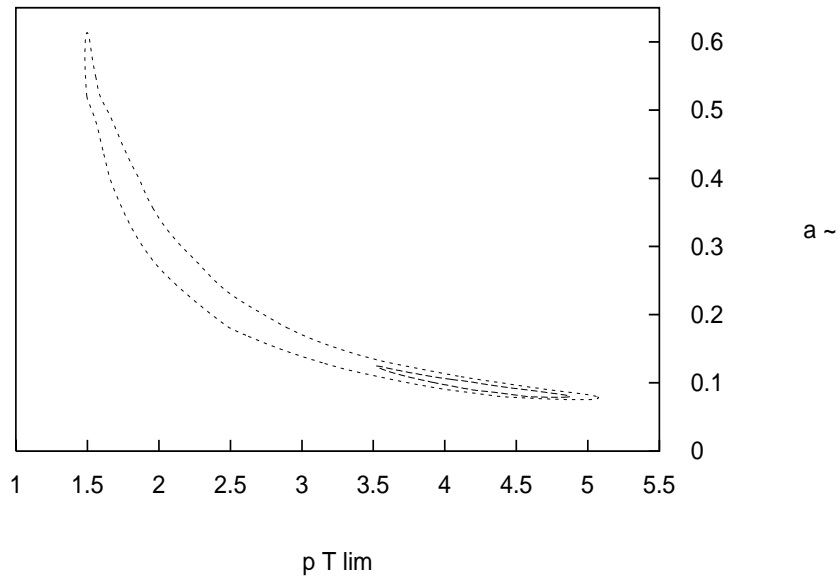,height=16cm,width=12cm,angle=270}}
\end{center}
\vspace{-1.5cm}
\caption{The contours of equal $\chi^2$   in the $\tilde{a}$, $\ptlim$ plane
for the KS $\pt$ space approach with the non-perturbative input
of the form (\ref{F:NP:pt},\ref{ptstar}). Both CDF and D0 data (with separate
normalization) for $\pt < 15$~GeV are used in the fit. The outer and inner
contours correspond to $\chi^2$/d.o.f. = 1 and 0.75 respectively. } 
\label{atildeptlim}
\end{figure}

  Generally we find that the best 
$\chi^2 /$d.o.f. value is obtained by  values of $\sqrt{1/\tilde{a}}$ and
 $\ptlim$ of order $3-4$~GeV. In this context the values proposed by
the EV collaboration $\tilde{a}=0.1$~GeV$^{-2}$, $\ptlim=4$~GeV provide one
of the best fits and describe the $Z$ data well. This
is also in agreement with the CDF and D0 analysis, cf.~\cite{CDF:Z,D0:Z}.
Furthermore we find that 
there is a wide range of strongly correlated values of larger $\tilde{a}$ and
smaller $\ptlim$ for which  $\chi^2 /$d.o.f. is only minimally worse.
This is illustrated in 
Fig.~\ref{atildeptlim}, which shows the equal $\chi^2$ contours in the 
 plane for $\chi^2$/d.o.f. = 1 and 0.75.

The correlation between  $\tilde{a}$ and $\ptlim$ is easy to understand.
Increasing $\ptlim$ corresponds to requiring the \nprt contribution to describe the 
data out to a larger value of $\pt$, and therefore a broader gaussian distribution
(equivalently, smaller  $\tilde{a}$) is required. The fact that the fit deteriorates sharply
as $\ptlim$ is made very small shows that (`frozen')
 perturbation theory alone cannot describe the data over the whole $\pt$ range.

Given the large variation in the allowed values of $\tilde{a}$ and $\ptlim$, it is difficult to
gauge the predictive power of these results, especially when one allows for
a possible additional $x$ and $Q$ dependence.   We also find that with the current
experimental data there is no need to introduce additional overall
smearing, as proposed in~\cite{EV}. A modification of $\tfnp$, such
as adding a linear term in the exponential or using a different freezing method,
does not significantly improve the fit either. 

\begin{figure}[hp]
\begin{center}
\mbox{\epsfig{figure=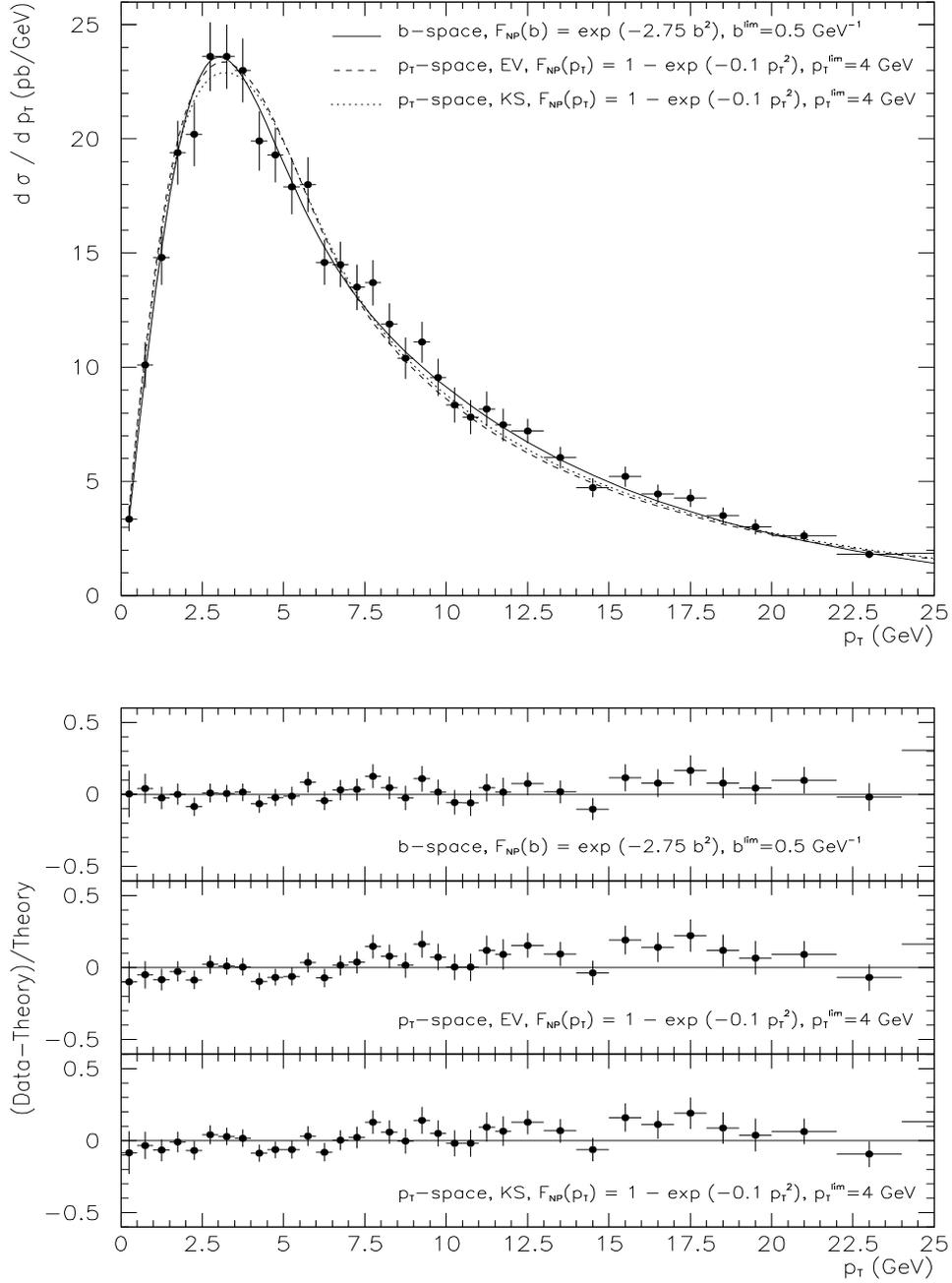,height=18cm,width=15cm}}
\end{center}
\caption{Comparison between CDF data on $Z$ production and theoretical predictions for the $b$
  space method, $\pt$ space method in the EV approach and in the KS
  approach. For the $b$ space method we use an effective gaussian form of
$F^{NP}$ as in~\cite{ERV}.} 
\label{CDFcompdatath}
\end{figure}

\begin{figure}[hp]
\begin{center}
\mbox{\epsfig{figure=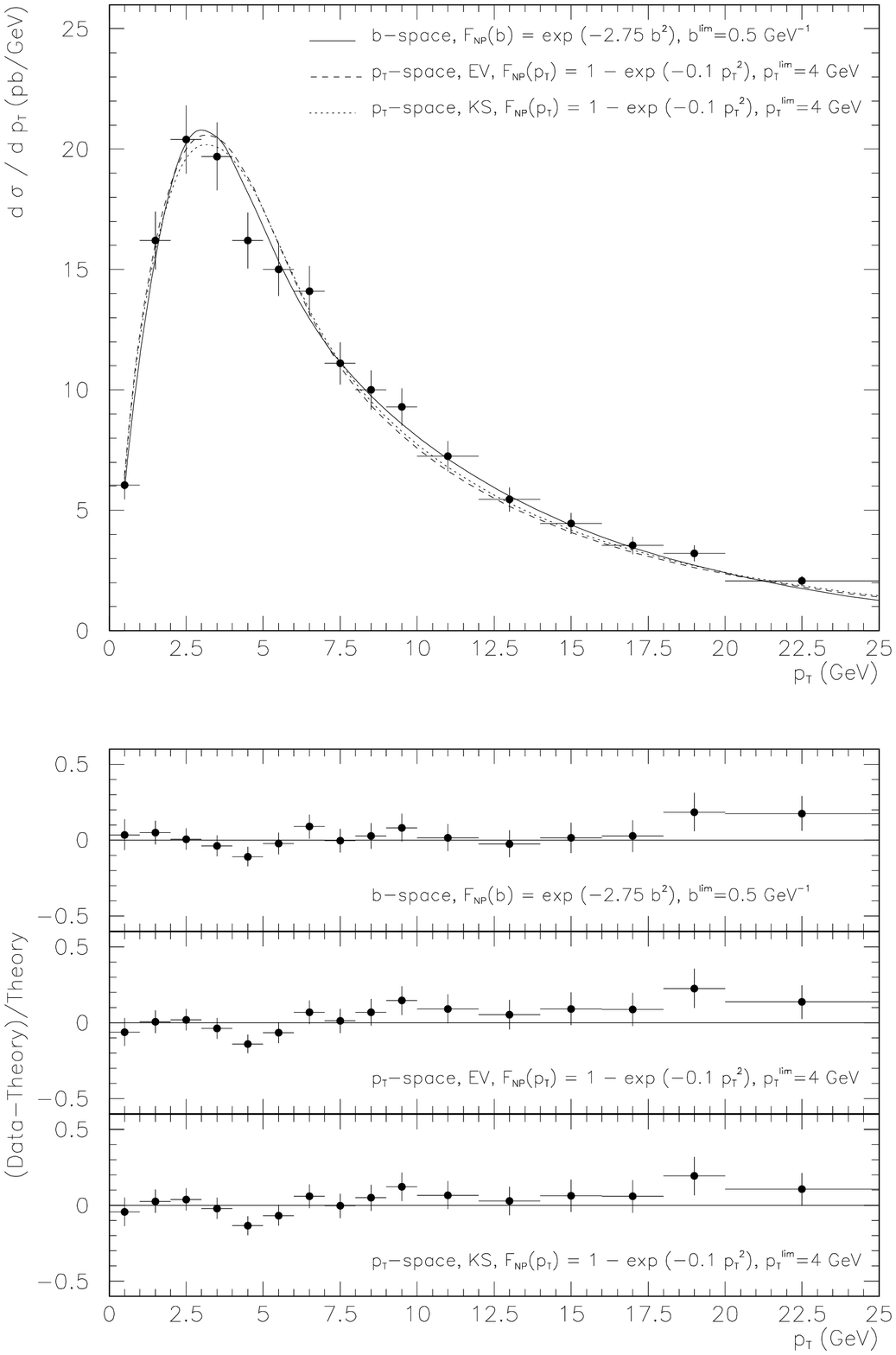,height=18cm,width=15cm}}
\end{center}
\caption{Comparison between D0 data on $Z$ production and theoretical predictions for the $b$
  space method, $\pt$ space method in the EV approach and in the KS
  approach. For the $b$ space method we use an effective gaussian form of
$F^{NP}$ as in~\cite{ERV}.}
\label{D0compdatath}
\end{figure}

\begin{figure}[hp]
\begin{center}
\mbox{\epsfig{figure=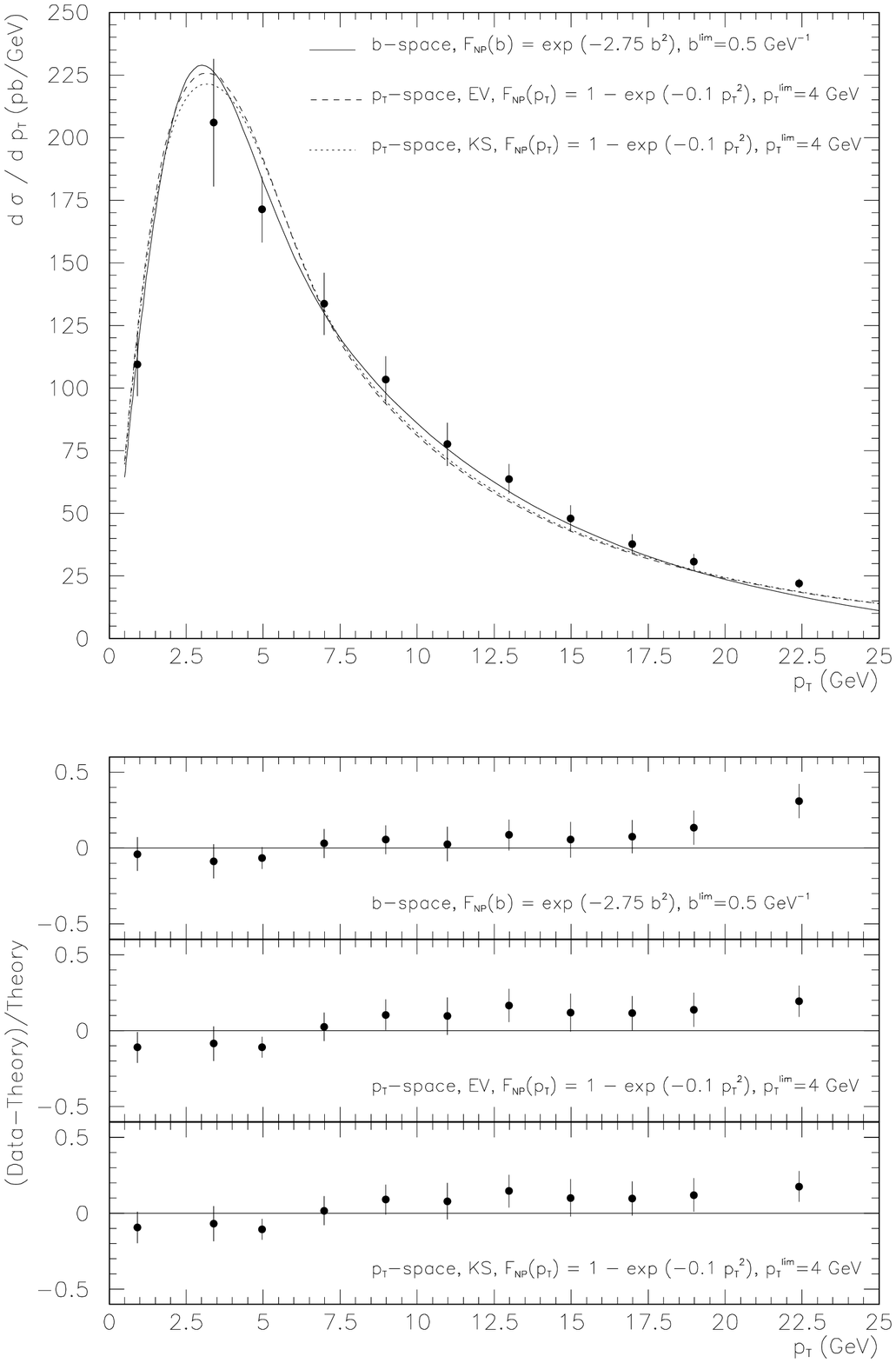,height=18cm,width=15cm}}
\end{center}
\caption{Comparison between D0 data on $W$ production and theoretical predictions for the $b$
  space method, $\pt$ space method in the EV approach and in the KS
  approach. For the $b$ space method we use an effective gaussian form of
$F^{NP}$ as in~\cite{ERV}.}
\label{WD0compdatath}
\end{figure}

In Figs.~\ref{CDFcompdatath},~\ref{D0compdatath},~\ref{WD0compdatath} we present a comparison between experimental data on $Z$ production
as measured by CDF and D0, $W$ production as measured by D0, and 
various  theoretical distributions  calculated using (i)
the $b$ space method, (ii) the EV $\pt$ space method,
 and (iii) the  KS $\pt$ space method. 
We observe good agreement between the data and the theoretical predictions for
all three methods, in the range of $\pt = 0 \sim 25$~GeV. In general, the $b$ 
space distribution is more `peaked' than the $\pt$ space equivalents. This
effect is, however, very susceptible to the choice of the \nprt function and
values of the \nprt parameters. The $b$ space distribution is
also higher in the intermediate range of $\pt= 10\sim 20$ GeV, where the
\nprt physics does not influence the resummed \prt result. In this region the 
KS distribution approximates the $b$ space result better than the
corresponding EV distribution. Given that the KS formalism resums more
towers of logarithms than the EV formalism, this is an expected result.
The increase of the cross section due to incorporating the 
fourth, NNNL, tower can be as big as  4\% for some values of $\pt$, both for
$W$ and $Z$ production \cite{KS}. 
 Interestingly we also observe a significant sensitivity to the value of $\as(M_Z)$ used
in the calculations.  A variation of $\as(M_Z)$ by $\pm 0.005$
around its average value, 0.1175, can cause, for some values of $\pt$, a more than $\pm 8$\%  change in the $Z$ $\pt$ distribution.


\begin{figure}[h]
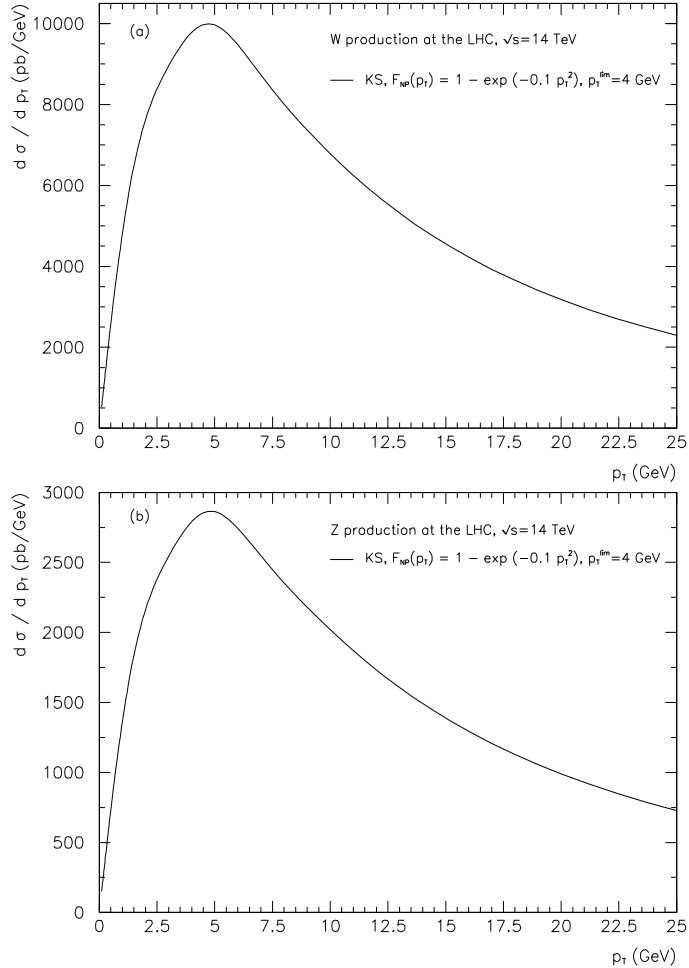

\begin{center}
\mbox{\epsfig{figure=wlhc.epsi,height=6.4cm,width=9cm}}
\mbox{\epsfig{figure=zlhc.epsi,height=6.4cm,width=9cm}}
\end{center}
\vspace{-.3cm}
\caption{$\pt$ space predictions (KS formalism) for the transverse momentum
  distribution at the LHC of: (a) $W$ boson, (b) $Z$ boson.} 
\label{WZLHC}
\vspace{-0.2cm}
\end{figure}

 The transverse momentum
distribution of $W$'s and $Z$'s at the LHC, predicted in the $\pt$ space formalism (KS), is
shown in Fig.~\ref{WZLHC}. The results agree with similar analyses performed
 using the $b$ space method~\cite{LesHouches}. For the
sake of this analysis we used the standard Tevatron values of the \nprt
parameters $\ta=0.1$ GeV$^{-2}$ and $\ptlim=4$~GeV in $\pt$ space. This may
prove to be a very unwise assumption if the \nprt parameterization does depend
 significantly on the partons momentum fractions, for example in the way it
was proposed for the $b$ space method~(\cite{LY}). However it does provide 
a useful benchmark and a reasonable `first guess'.

\vspace{-0.3cm}
\section{Summary}

We have applied the KS resummation technique in $\pt$ space,
developed in~\cite{KS}, to the hadronic production of vector
bosons. At the hadron level our approach retains the potential of the full
resummation of the first four towers of logarithms. We also allow for a non-perturbative
contribution, with a smooth interpolation between the perturbative and 
non-perturbative regimes at small $\pt$.
Our  numerical results generally show good
agreement with  recent data on $W$ and $Z$ boson production from the
Tevatron collider.

For the resummed part of the $ d \sg / d \pt$
distribution we observe rather weak dependence on the renormalization scale,
but some sensitivity to the value of $\as(M_Z)$. 
However the non-perturbative contribution, which at present must be determined
from fits to data,  is not well determined.  We have not attempted to estimate the full error
on the \nprt contribution. Rather, we showed that a simple Gaussian form, with reasonable
values of the parameters, gives an acceptable fit.
The width of the gaussian and the transition point between the 
perturbative and non-perturbative regions are, however, strongly correlated.
This, combined with the lack of knowledge of the $x$ and $Q$ dependence
of the \nprt parameters, makes it difficult to formulate precise predictions
for the corresponding distributions at the LHC.

\vspace{.2cm}
\noindent
{\bf Acknowledgements}:
This work was supported in part by the EU Fourth Framework Programme
`Training and Mobility of Researchers', Network `Quantum Chromodynamics
and the Deep Structure of Elementary Particles', contract FMRX-CT98-0194
(DG 12 - MIHT).


\end{document}